\documentclass[aps,prd,twocolumn,preprintnumbers,amsmath,amssymb,
superscriptaddress,showpacs,floatfix]{revtex4}

\usepackage{graphicx}
\usepackage{dcolumn}
\usepackage{bm}
\usepackage{epsfig}
\usepackage{graphics}
\usepackage{latexsym}

\def\beq{\begin{equation}}
\def\eeq{\end{equation}}
\def\bea{\begin{eqnarray}}
\def\eea{\end{eqnarray}}
\newcommand{\beqs}{\begin{subequations}}
\newcommand{\eeqs}{\end{subequations}}

\newcommand{\cref}[1]{Ref.~\cite{#1}}

\newcommand{\Tr}{\mathsf{Tr}}

\newcommand{\hh}{{\ensuremath{I{\kern-2.6pt h}}}}
\newcommand{\bhh}{{\ensuremath{\bar{I{\kern-2.6pt h}}}}}

\begin{document}

\preprint{UT-STPD-16/01}

\title{Axion Model with Intermediate Scale 
Fermionic Dark Matter}

\author{G. Lazarides}
\email{lazaride@eng.auth.gr} \affiliation{School of Electrical and
Computer Engineering, Faculty of Engineering, Aristotle University
of Thessaloniki, Thessaloniki 54124, Greece}
\author{Q. Shafi}
\email{shafi@bartol.udel.edu}
\affiliation{Bartol Research Institute, Department of Physics and 
Astronomy, University of Delaware, Newark, DE 19716, USA}

\date{\today}

\begin{abstract}
We investigate a non-supersymmetric $SO(10)\times U(1)_{\rm PQ}$ 
axion model in which the spontaneous breaking of $U(1)_{\rm PQ}$ 
occurs after inflation, and the axion domain wall problem is 
resolved by employing the Lazarides-Shafi 
mechanism. This requires the introduction of two fermion 
10-plets, such that the surviving discrete symmetry from the
explicit $U(1)_{\rm PQ}$ breaking by QCD instantons is reduced 
from $Z_{12}$ to $Z_4$, where $Z_4$ coincides with the center 
of $SO(10)$ (more precisely $Spin(10)$). An unbroken $Z_2$ 
subgroup of $Z_4$ yields 
intermediate scale topologically stable strings, as well as a 
stable electroweak doublet non-thermal dark matter candidate 
from the fermion 10-plets with mass comparable to or somewhat 
smaller than the axion decay constant $f_{\rm a}$. We present 
an explicit realization with inflation taken into account and 
which also incorporates non-thermal leptogenesis. The fermion 
dark matter mass lies in the $3\times 10^{8}-10^{10}~{\rm GeV}$ 
range and its contribution to the relic dark matter abundance 
can be comparable to that from the axion.
\end{abstract}

\pacs{12.60.Jv} 
\maketitle

\section{Introduction}
\label{sec:intro}

An elegant resolution of the strong CP problem is provided 
by the Peccei-Quinn (PQ) mechanism \cite{PecceiQuinn}, which 
also predicts the existence of axion, a compelling dark 
matter (DM) candidate \cite{axionDM}.
In a relatively simple but realistic non-supersymmetric 
$SO(10)\times U(1)_{\rm PQ}$ axion  model constructed 
sometime ago \cite{holman}, the well-known axion domain wall 
problem \cite{sikivie} is resolved without invoking inflation 
through implementation of the Lazarides-Shafi mechanism 
\cite{axionwalls}. This is achieved by introducing two 
10-plets of fermions carrying appropriate charges under the 
$U(1)_{\rm PQ}$ symmetry \cite{PecceiQuinn}, such that the 
residual discrete symmetry from the explicit $U(1)_{\rm PQ}$ 
breaking by the QCD instantons coincides with the center of 
$SO(10)$ (more precisely $Spin(10)$). This construction 
therefore allows one 
to implement the spontaneous breaking of $U(1)_{\rm PQ}$ in a 
post-inflationary phase. Among other things the model evades 
some thorny issues such as the isocurvature problem \cite{iso} 
that appears if $U(1)_{\rm PQ}$ breaks during inflation. 
For some recent papers on $U(1)_{\rm PQ}$ breaking during 
inflation see Ref.~\cite{PQinf}.

The model in Ref.~\cite{holman} employs only tensor 
representations in order to break $SO(10)$ via at least one 
intermediate stage to the Standard Model (SM) gauge group, and 
subsequently to $SU(3)_c\times U(1)_{\rm em}$. This means 
that a discrete $Z_2$ subgroup of $Z_4$, the center of $SO(10)$, 
remains unbroken \cite{kibble}, which has some important 
consequences. Firstly, there exist non-superconducting 
topologically stable cosmic 
strings which, depending on their mass scale, may survive 
inflation. Secondly, in the framework of supersymmetry, the 
leftover $Z_2$ symmetry is precisely `matter' parity which, 
among other things, ensures that the lightest supersymmetric 
particle is stable. Finally, in the non-supersymmetric 
$SO(10)\times U(1)_{\rm PQ}$ model under discussion, the 
presence of the unbroken $Z_2$ symmetry \cite{kibble}, offers 
a new DM particle candidate arising from the 10-plet fermions 
whose mass is related to the intermediate scale of the PQ 
symmetry breaking. For an 
earlier example of intermediate scale DM particles coexisting 
with axions see Ref.~\cite{intermediate}; for superheavy DM 
(wimpzillas) see Ref.~\cite{wimpzilla}. It should be noted 
that this unbroken $Z_2$ symmetry \cite{kibble} has been 
employed \cite{other,ringwald}, in recent years, to guarantee 
DM stability in non-supersymmetric $SO(10)$ models. 
(Ref.~\cite{ringwald} also contains a brief 
discussion of our $SO(10)\times U(1)_{\rm PQ}$ model.)  
The proposed scenarios, though, are generally very different 
from ours with the DM particle masses in the TeV range or so.   

The plan of this paper is as follows. In Sec.~\ref{sec:model}, 
we present the salient feature of our $SO(10)$ model coupled 
to the inflationary scheme of Ref.~\cite{Shafi:1983bd}, where 
the inflaton is a gauge singlet scalar field with a
Coleman-Weinberg potential. In Sec.~\ref{sec:reheat}, we 
describe the reheating process following the end 
of inflation and, in Sec.~\ref{sec:DM}, we examine the 
possibility that the fermion 10-plets provide a novel 
DM candidate with intermediate scale mass. 
Sec.~\ref{sec:lepto} is devoted to 
the discussion of the generation of the baryon asymmetry of 
the Universe (BAU) via leptogenesis. Our results are 
summarized in Sec.~\ref{sec:concl}. 

\section{The model}
\label{sec:model}

We consider the non-supersymmetric $SO(10)$ grand unified 
theory (GUT) model introduced in Ref.~\cite{holman}. The
model contains a global anomalous PQ 
symmetry ${\rm U(1)}_{\rm PQ}$ \cite{PecceiQuinn}. The 
fermion content consists of the following $SO(10)$ 
multiplets,
\beq
\psi_{16}^{(i)}\, (i=1,2,3), \quad \psi_{10}^{(\alpha)}\, 
(\alpha=1,2),
\eeq   
where the subscripts denote the dimension of the 
representation. The PQ charges of the $\psi_{16}^{(i)}$'s 
and $\psi_{10}^{(\alpha)}$'s are 1 and -2 respectively. 
The scalar Higgs fields are
\beq
\varphi_{210}\, (0),\quad \varphi_{126}\, (2), \quad
\varphi_{45}\, (4), \quad \varphi_{10} \,(-2)
\eeq
with the PQ charges $Q_{\rm PQ}$ indicated in parentheses. 
Note that the discrete subgroup of $U(1)_{\rm PQ}$ left 
unbroken by the instantons is $Z_4$ and coincides with the 
center of $SO(10)$. Consequently, the axion domain wall 
problem \cite{sikivie} does not arise \cite{axionwalls} 
even if the PQ symmetry is broken after the end of 
inflation. The allowed Yukawa couplings are
\beq
\psi_{16}\psi_{16}\varphi_{10},\quad \psi_{16}\psi_{16}
\varphi_{126}^\dagger,\quad \psi_{10}\psi_{10}\varphi_{45},
\label{Ycouplings}
\eeq
while the Higgs couplings include 
\beq
\varphi_{210}\varphi_{126}^\dagger\varphi_{126}^\dagger
\varphi_{45}, \quad \varphi_{210}\varphi_{126}^\dagger
\varphi_{10}\varphi_{45}, \quad \varphi_{210}\varphi_{126}
\varphi_{10}.
\label{Hcouplings}
\eeq

The $SO(10)$ symmetry breaking to the SM in a 
non-supersymmetric setting usually proceeds via one or 
more intermediate stages \cite{joydeep}.
For definiteness, we assume the following symmetry 
breaking chain:
\begin{gather}
SO(10)\times U(1)_{\rm PQ}\overset{\varphi_{210}}
{\underset{M_{\rm G}}{\longrightarrow}} 
\nonumber\\
SU(3)_c\times SU(2)_L\times SU(2)_R\times U(1)_{B-L}
\times U(1)_{\rm PQ}
\overset{\varphi_{126}}{\underset{M_{\rm I}}
{\longrightarrow}} 
\nonumber\\
SU(3)_c\times SU(2)_L\times U(1)_Y\times Z_2
\times U(1)_{\rm PQ}^\prime
\overset{\varphi_{45}}{\underset{f_{\rm a}}
{\longrightarrow}}
\nonumber\\
SU(3)_c\times SU(2)_L\times U(1)_Y\times Z_2
\overset{\varphi_{10},\varphi_{126}}{\underset
{M_{\rm W}}
{\longrightarrow}}
\nonumber\\
SU(3)_c\times U(1)_{\rm em}\times Z_2, 
\label{symbreak}
\end{gather}
where the Higgs fields implementing the breaking chain 
and the corresponding scales are indicated. The first
breaking in Eq.~(\ref{symbreak}) is achieved by the 
vacuum expectation value (VEV) of $\varphi_{210}$ along 
its (1,1,1) and (15,1,1) components with respect to 
the subgroup $G_{\rm PS}=SU(4)_c\times SU(2)_L\times 
SU(2)_R$ \cite{PatiSalam}. The PQ symmetry is left 
unbroken, but superheavy magnetic monopoles are created 
during this breaking. As we shall see they are inflated 
away. 

The next breaking at the intermediate scale $M_{\rm I}$ 
is achieved by the VEV of $\varphi_{126}$ along its 
($\overline{10}$,1,3) component. This leaves the $Z_2$ 
subgroup of $Z_4$ unbroken, leading to the 
formation of intermediate scale topologically stable 
$Z_2$ cosmic strings \cite{kibble}, which are not 
\cite{topdef} superconducting \cite{supercond}.
The PQ symmetry, however, is not broken at this stage. 
It is merely rotated to $U(1)_{\rm PQ}^\prime$, with 
$Q_{\rm PQ}^\prime=(5Q_{\rm PQ}+\chi)/4$, where 
$\chi=-3(B-L)+4T_R^3$ is the generator of the $U(1)_\chi$ 
subgroup of $SO(10)$ which is not contained in $SU(5)$. 
In order to see this, note that the $\chi$ charges of the 
usual quarks and leptons in $\psi_{16}$ are  
$\nu^c (-5),~u^c,q,e^c (-1),~d^c,l (3)$, and the 
$\chi$ charges of the color anti-triplets $D^c$ 
(triplets $\overline{D^c}$) and $SU(2)_L$ doublets $L$
(anti-doublets $\overline{L}$) in $\psi_{10}$ are -2 (2).
The $\nu^c\nu^c$-type component of $\varphi_{126}$ 
has charges $Q_{\rm PQ}=2$, $\chi=-10$, as one can see 
from the 
$\chi$ charge of $\nu^c$. So the unbroken PQ symmetry 
is indeed $U(1)_{\rm PQ}^\prime$ with the 
$Q_{\rm PQ}^\prime$ charges of the various fields given by 
$\nu^c (0),~u^c,q,e^c (1),~d^c,l (2),
~D^c,L (-3),~ \overline{D^c},\overline{L} (-2)$. The 
discrete subgroup $Z_N$ of $U(1)_{\rm PQ}^\prime$ left 
unbroken by the QCD instantons is $Z_5$, since from $q,u^c$ 
we have $3\times 3$ color triplets and anti-triplets with 
$Q_{\rm PQ}^\prime=1$, from $d^c$ three color anti-triplets 
with $Q_{\rm PQ}^{\prime}=2$, from $D^c$ two anti-triplets 
with $Q_{\rm PQ}^{\prime}=-3$, and from $\overline{D^c}$ 
two triplets with $Q_{\rm PQ}^{\prime}=-2$. Therefore, 
altogether $N=3\times 3+3\times 2-2\times 3-2\times 2=5$. 
This $Z_5$ coincides with the $Z_5$ subgroup of $U(1)_Y$ 
generated by $\exp[(i2\pi/5)6Y]$ and, therefore, is not a 
genuine discrete symmetry. It is instructive to further 
rotate the PQ symmetry to $U(1)_{\rm PQ}^{\prime\prime}$ 
with $Q_{\rm PQ}^{\prime\prime}=(-6Y+Q_{\rm PQ}^{\prime})/5$. 
The $Q_{\rm PQ}^{\prime\prime}$ charges of the various fields 
are given by $\nu^c,q,d^c,L,\overline{D^c} (0),~u^c,l(1),
~e^c,D^c,\overline{L}(-1)$. In this case, $N=3-2=1$ and 
therefore only the identity element of $U(1)_{\rm PQ}
^{\prime\prime}$ is left unbroken by the QCD instanton 
effects. 

The PQ breaking at a scale $f_{\rm a}$ (the axion decay 
constant), which can be close to $M_{\rm I}$, is achieved 
by the VEVs of $\varphi_{45}$ along its (15,1,1) and (1,1,3) 
components. These VEVs have $Q_{\rm PQ}^{\prime\prime}=1$ 
since they couple with $D^c \overline{D^c}$ and 
$L\overline{L}$ respectively. Consequently, they break 
spontaneously $U(1)_{\rm PQ}^{\prime\prime}$ to its identity 
element. The axion strings from the $U(1)_{\rm PQ}^{\prime
\prime}$ breaking acquire, at 
the QCD transition, just one axion domain wall (for walls
bounded by strings see Ref.~\cite{wallsbounded}) and, thus, 
the string-wall network decays \cite{axionwalls}. Therefore, 
the troublesome axion domain wall cosmological problem 
\cite{sikivie} is avoided.
Finally, the electroweak symmetry breaking is achieved by the 
VEV of a linear combination of the (1,2,2) component of 
$\varphi_{10}$ and the (15,2,2) component of $\varphi_{126}$. 
Note that the $Z_2$ subgroup of $U(1)_{B-L}$ is neither in 
$U(1)_{\rm PQ}^{\prime\prime}$ nor the SM gauge group. It 
is a genuine discrete symmetry which is left unbroken by all 
the VEVs and, thus, the corresponding intermediate scale 
cosmic strings \cite{kibble} can survive until the present 
time. These strings are not superconducting \cite{topdef}.

Next, we merge our $SO(10)$ GUT model with the inflationary 
model of Refs.~\cite{Shafi:1983bd,extended}, where inflation 
is driven by a $SO(10)\times U(1)_{\rm PQ}$ singlet real 
scalar field $\phi$ with a Coleman-Weinberg potential and 
with minimal coupling to gravity:
\beq
V(\phi)=A\phi^4\left(\ln\left(\frac{\phi}{M}\right)-
\frac{1}{4}\right)+V_0.
\label{CW}
\eeq 
Here $M$ is the VEV of $\phi$, $V_0=AM^4/4$, and we employ 
the cutoff regularization. This model predicts that the 
tensor-to-scalar ratio $r\gtrsim 0.01$ \cite{Shafi:2006cs}. 
The important couplings which induce 
the VEVs of the various scalar fields $\varphi_\theta$ 
($\theta=210,126,45,10$) as the inflaton acquires its final 
VEV, starting below $M$, are 
\beq
-\frac{c_{\theta}}{4}\phi^2\varphi_{\theta}^2+
\frac{\alpha_{\theta}}{4}\varphi_{\theta}^4,
\label{ctheta}
\eeq
where the real canonically normalized component of a
scalar field which acquires a VEV is represented by the 
same symbol as the field. The VEVs and the masses of 
the scalar fields are 
\begin{equation}
\langle\varphi_{\theta}\rangle^2=\frac{c_{\theta}M^2}
{2\alpha_{\theta}},\quad
m_{\theta}^2=c_{\theta}M^2.
\label{VEV}
\end{equation}

To be more specific, we will consider  
a particular viable realization of this inflationary 
scenario which appears in the fourth line of Table 4 in 
Ref.~\cite{okada}. In this case, the inflationary scale 
$V_0^{1/4}\simeq 1.75\times 10^{16}~{\rm GeV}$, $A\simeq 
1.43\times 10^{-14}$, and the VEV of the inflaton $M
\simeq 7.17\times 10^{19}~{\rm GeV}$. Following
Ref.~\cite{Shafi:1983bd}, we evaluate the coefficient 
$A$ of the Coleman-Weinberg potential in Eq.~(\ref{CW}) 
from the
radiative corrections arising from the term $(-1/4)
c_{210}\phi^2\varphi_{210}^2$, which are the dominant 
ones. We find
\beq
A=\frac{210}{64\pi^2}\,c_{210}^2,
\eeq
which gives $c_{210}\simeq 2.07\times 10^{-7}$ and, from 
Eq.~(\ref{VEV}) with $\alpha_{210}=1/2$,
$M_{\rm G}\equiv\langle\varphi_{210}\rangle\simeq 3.27
\times 10^{16}~{\rm GeV}$. From Eq.~(\ref{CW}), we find the 
inflaton mass $m_\phi\simeq 1.7\times 10^{13}~{\rm GeV}$.

\begin{figure}[t]
\centerline{\epsfig{file=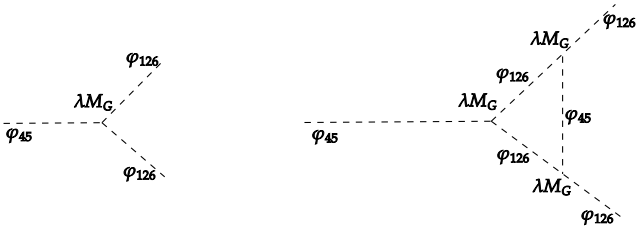,width=8.4cm}}
\caption{Diagrams for the decay of the (1,1,3) component 
of $\varphi_{45}$ to a pair of electroweak Higgs fields 
contained in the (15,2,2) component of $\varphi_{126}$. 
Dashed lines represent scalar bosons.}
\label{fig:trilinear}
\end{figure}

It is important to note that the dimensionless coupling
constants of the two scalar quartic interactions in 
Eq.~(\ref{Hcouplings}) should be suppressed by at least
$m_{45}/M_{\rm G}$ for the model to remain
perturbative. One can see this by considering, for 
example, the first coupling in this equation with 
dimensionless coefficient $\lambda$ and replace
$\varphi_{210}$ by its VEV along its (1,1,1) component. 
We then obtain a trilinear scalar coupling with 
coefficient of order $\lambda M_{\rm G}$. The 
(1,1,3) component of $\varphi_{45}$ can decay via this 
trilinear coupling to a pair of electroweak Higgs 
fields contained in the (15,2,2) component of 
$\varphi_{126}$ (see Fig.~\ref{fig:trilinear}). 
The vertex diagram in this figure can be radiatively 
corrected by inserting the same trilinear coupling on 
the two external $\varphi_{126}$ lines with an exchange 
between them of the (1,1,3) component of $\varphi_{45}$. 
The radiative correction acquires a factor of order
$\lambda^2 M_{\rm G}^2/m_{45}^2$ relative to the
tree diagram. The requirement of perturbativity then 
implies that $\lambda\lesssim m_{45}/M_{\rm G}$.

Potentially dangerous trilinear scalar couplings can 
also arise from the first term in Eq.~(\ref{ctheta}).
For instance, setting $\phi=M+\delta\phi$, we obtain 
the trilinear scalar coupling $(-1/2)c_{45} M \phi
\varphi_{45}^2$. Considering the decay of the inflaton 
via this coupling into a pair of $\varphi_{45}$'s and 
repeating the argument of the previous paragraph, we 
find that the radiative correction acquires a factor 
of order $c_{45}^2M^2/m_\phi^2=(m_{45}^2/M m_\phi)^2$ 
relative to the tree diagram. However, this factor is 
much smaller than unity as we will see below.
This conclusion remains true even if we replace 
$\varphi_{45}$ by $\varphi_{126}$ or $\varphi_{10}$.
In the case of $\varphi_{210}$, this danger is not 
encountered since the inflaton mass is much smaller 
than $M_{\rm G}$ (see below), and the inflaton decay 
into two $\varphi_{210}$'s is kinematically blocked. 
One could also insert $\varphi_{\theta}=\langle
\varphi_{\theta}\rangle+\delta\varphi_{\theta}$ in 
the first term in Eq.~(\ref{ctheta}) to obtain the 
trilinear coupling
$(-1/2)c_{\theta}\phi^2\langle\varphi_{\theta}\rangle
\delta\varphi_{\theta}$. The only field which is
kinematically allowed to decay to two inflatons via
this coupling is $\delta\varphi_{210}$. Perturbativity 
then requires that $c_{210}=M_{\rm G}^2/M^2\ll 1$,
which is well satisfied. Finally, we should mention 
that, substituting $\phi=M+\delta\phi$ and 
$\varphi_\theta=\langle\varphi_\theta\rangle+
\delta\varphi_\theta$ in the first term in 
Eq.~(\ref{ctheta}), we obtain a bilinear mixing term 
between $\delta\phi$ and $\delta\varphi_\theta$: 
\beq
-Mc_\theta\langle\varphi_\theta\rangle\delta\phi\delta
\varphi_\theta.
\label{bilinear}
\eeq
As we will see later, these scalar couplings will be 
important in the inflaton decay. 

\begin{figure}[t]
\centerline{\epsfig{file=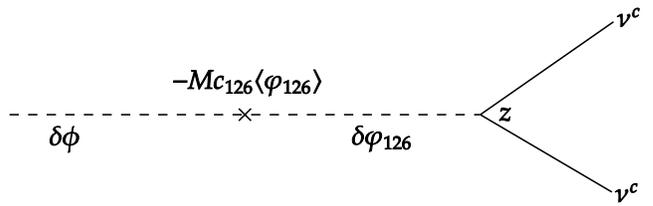,width=8.4cm}}
\caption{Diagram for the inflaton decay to a pair of 
right-handed neutrinos $\nu^c$. Solid lines represent 
fermions, while dashed lines represent scalar bosons.}
\label{fig:infdecaytonuc}
\end{figure}

\section{Reheating}
\label{sec:reheat} 

With the quartic scalar couplings in Eq.~(\ref{Hcouplings}) 
adequately suppressed as discussed above, the main 
decay mode of the inflaton $\delta\phi$ will be to a 
pair of right-handed neutrinos $\nu^c$ via the diagram 
depicted in 
Fig.~\ref{fig:infdecaytonuc}. This decay has to 
be out of equilibrium and to the second heaviest 
right-handed neutrino $\nu^c_2$ in order to provide the 
possibility of generating the BAU via the scenario of 
non-thermal \cite{nonthermalepto,vlachos} leptogenesis 
\cite{thermallepto} (for a review see 
Ref.~\cite{pedestrians}). The cross sign in 
Fig.~\ref{fig:infdecaytonuc} represents the coefficient 
$-Mc_{126}\langle\varphi_{126}\rangle$ of the effective 
bilinear coupling between $\delta\phi$ and 
$\delta\varphi_{126}$ 
(see Eq.~(\ref{bilinear})), $z$ is the Yukawa coupling
constant of $\varphi_{126}$ with $\nu^c_2$, and 
$\langle\varphi_{126}\rangle$ as well as the propagating 
$\delta\varphi_{126}$ lie along the $\nu^c\nu^c$-type 
component of $\varphi_{126}$. For definiteness, we take 
$\alpha_{126}=1/2$. Eq.~(\ref{VEV}) then implies that 
\begin{equation}
\langle\varphi_{126}\rangle=m_{126},\quad c_{126}=
\left(\frac{m_{126}}{M}\right)^2.
\end{equation}
An important requirement for the mass $M_2$ of $\nu^c_2$
is that 
\begin{equation}
M_2\equiv M_{\nu^c_2}= \langle\varphi_{126}\rangle z
\leq \frac{1}{2}m_\phi.
\label{M2}
\end{equation}
This guarantees that the decay $\delta\phi\to\nu^c_2
\nu^c_2$ is kinematically possible.

For $m_{126}\leq m_\phi$, the propagator of $\delta
\varphi_{126}$ is dominated by the mass of the 
inflaton and the decay width is given by
\begin{gather}
\Gamma_{\phi\to\nu^c}\simeq\frac{1}{16\pi}\left(\frac{Mc_{126}
\langle\varphi_{126}\rangle z}{m_\phi^2}\right)^2 m_\phi=
\nonumber\\
\frac{1}{16\pi}\left(\frac{m_{126}^2 M_2}{Mm_\phi^2}
\right)^2 
m_\phi.
\end{gather} 
In the following we will not consider loop corrections in 
the estimates for decay widths. We make the self-consistent
assumption here and in what follows that the trilinear 
couplings are sufficiently small for this to hold.
Saturating the inequality in Eq.~(\ref{M2}) and taking 
$m_{126}=m_\phi$, we find the maximal allowed 
$\Gamma_{\phi\to\nu^c}$ in this case:
\begin{equation}
\Gamma_{\phi\to\nu^c}\simeq\frac{1}{16\pi}\left(\frac{m_\phi}
{2M}\right)^2 m_\phi\simeq 4.84\times 10^{-3}~{\rm GeV}.
\label{Gnuc}
\end{equation}
Note that phase space factors are not taken into account 
here. The impact is minimal in our estimate for the reheat 
temperature $T_{\rm r}$ below.
For $m_{126}\geq m_\phi$, on the other hand, the propagator 
should be replaced by $1/m_{126}^2$, which gives
\begin{equation}
\Gamma_{\phi\to\nu^c}\simeq\frac{1}{16\pi}\left(\frac{Mc_{126}
\langle\varphi_{126}\rangle z}{m_{126}^2}\right)^2 m_\phi=
\frac{1}{16\pi}\left(\frac {M_2}{M}\right)^2 m_\phi.
\label{Gamma}
\end{equation}
This is again maximized at the same value of $M_2$ as in the 
previous case. The corresponding (maximal) reheat 
temperature turns out to be 
\begin{equation}
T_{\rm r}\simeq\left(\frac{45}{2\pi^2g_*}\right)^{\frac{1}{4}}
(\Gamma_{\phi\to\nu^c}\,m_{\rm P})^{\frac{1}{2}}\simeq 4.15
\times 10^{7}~{\rm GeV},
\label{reheat} 
\end{equation}
for an effective number of degrees of freedom 
$g_*=106.75$ 
corresponding to the SM spectrum ($m_{\rm P}$ is
the reduced Planck mass). Note that 
the values of $\Gamma_{\phi\to\nu^c}$ in Eq.~(\ref{Gnuc}) and
$T_{\rm r}$ in Eq.~(\ref{reheat}) are independent of
$\langle\varphi_{126}\rangle=m_{126}$ provided that 
$\langle\varphi_{126}\rangle\geq m_\phi$. Also, the decay of 
right-handed neutrinos to SM particles is rapid enough to 
justify the formula in Eq.~(\ref{reheat}).

\begin{figure}[t]
\centerline{\epsfig{file=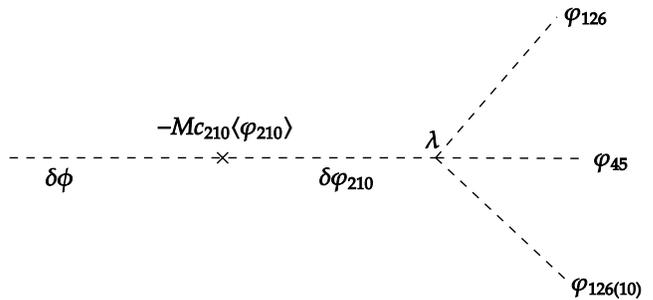,width=8.4cm}}
\caption{Diagram for the inflaton decay to a pair of 
electroweak Higgs fields and a SM singlet
scalar field. The conventions are as in 
Fig.~\ref{fig:trilinear}.}
\label{fig:infdecaytoh}
\end{figure}

One could alternatively consider the decay of the 
inflaton to a pair of electroweak Higgs fields and a 
SM singlet 
scalar field from $\varphi_{45}$. The relevant diagram 
is depicted in Fig.~\ref{fig:infdecaytoh} and uses the 
bilinear coupling between $\delta\phi$ 
and $\delta\varphi_{210}$. The electroweak Higgs fields 
are contained in the $\varphi_{126}$ or $\varphi_{10}$ 
external lines. The decay width is estimated to be
\beq
\Gamma_{\phi\to h}\simeq\frac{1}{192\pi^3}\left(\frac
{M c_{210}\langle\varphi_{210}\rangle\lambda}{m_{210}^2}
\right)^2m_\phi\simeq 595\, \lambda^2~{\rm GeV}.
\eeq 
Here $\lambda$ is the typical coefficient of the quartic 
scalar couplings in Eq.~(\ref{Hcouplings}), which, as we 
have shown, should not exceed $m_{45}/M_{\rm G}$. As we 
will see later $m_{45}\sim 10^{10}-10^{11}~{\rm GeV}$, 
which implies that $\Gamma_{\phi\to h}\lesssim 5.6\times 
10^{-9}~{\rm GeV}\ll \Gamma_{\phi\to\nu^c}$. So this 
decay mode is much less important than the decay into 
$\nu^c$'s. Note that $M_2= 
m_\phi/2\simeq 8.5\times 10^{12}~{\rm GeV}\gg T_{\rm r}$, 
and thus the out-of-equilibrium condition for its 
subsequent decay 
is well satisfied. It is finally important to keep in mind 
that we can consider smaller values of $M_2$ with any value 
of $m_{126}\geq m_\phi$. However, this will reduce 
$T_{\rm r}$ by the same proportion. Finally, we have 
checked that parametric resonance is not important in our 
scenario.

\section{Intermediate Scale Dark Matter}
\label{sec:DM}

We will now examine the possibility that the two neutral
Dirac fermions contained in $\psi_{10}^{(\alpha)}$ 
($\alpha=1,2$) can provide a new DM particle whose mass 
is of intermediate scale. The fermion
$\psi_{10}^{(\alpha)}$'s are the 
only fields in the model that are odd under an unbroken 
$Z_2$ symmetry, arising as a combination of 
the unbroken $Z_2$ subgroup of the $Z_4$ center of 
$SO(10)$ and the $Z_2$ fermion number symmetry, under 
which all fermions are odd. Thus, the lightest components 
of the $\psi_{10}^{(\alpha)}$'s, which 
are assumed to be their neutral components, cannot decay, 
but can only annihilate in pairs, which makes them 
potentially viable DM candidates. Note that each of the 
color triplets and anti-triplets in the two fermion 
10-plets can decay into the $SU(2)_L$ doublet in the same 
10-plet plus SM particles through the exchange of a 
superheavy gauge boson. We estimate their lifetime to be 
$\sim 10^{-5}~{\rm sec}$, and so these decays take 
place well before Big Bang Nucleosynthesis -- compare with 
the last paper in Ref.~\cite{other}. Also, each of the 
charged members of the $SU(2)_L$ doublets in the fermion 
10-plets can decay into the neutral member of the same 
doublet plus SM particles through the exchange of a 
$W^\pm_L$ gauge boson. Even a moderate mass splitting 
between the members of these doublets induced by loops of 
SM gauge bosons is enough to insure that these decays are 
very rapid -- see second paper in Ref.~\cite{witten}.    

The (1,2,2) components of the two $\psi_{10}$'s 
can be written as
\beq
H_{(\alpha)}=
\begin{pmatrix}
\bar{N}_{(\alpha)}& E_{(\alpha)}^-
\\
&
\\
E_{(\alpha)}^+ & N_{(\alpha)}
\end{pmatrix}, \quad \alpha=1,2, 
\eeq
where $\bar{N}_{(\alpha)}$ and $N_{(\alpha)}$ are electrically 
neutral. The field $\varphi_{45}$ is antisymmetric and thus
couples only with $\psi_{10}^{(1)}\psi_{10}^{(2)}$. 
Consequently, the coupling of its (1,1,3) component with the 
$H_{(\alpha)}$'s yields the mass term
\beq
y\langle\varphi_{45}\rangle \Tr\left(H_{(1)}\epsilon\sigma_3
\epsilon\tilde{H}_{(2)}\right)+{\rm h.c.},
\eeq
where $y$ is the Yukawa coupling constant, $\langle
\varphi_{45}\rangle$ will, from now on, represent the VEV 
of the (1,1,3) component of $\varphi_{45}$, $\epsilon$ is 
the antisymmetric $2\times 2$ matrix with $\epsilon_{12}=1$, 
and tilde denotes transposition. This, in turn, yields the 
mass term
\beq
y\langle\varphi_{45}\rangle\left(\bar{N}_{(1)}N_{(2)}
-N_{(1)}\bar{N}_{(2)}\right)+{\rm h.c.}
\eeq 
for the neutral components. At tree level we obtain two 
neutral Dirac fermions of equal mass $m_{\rm DM}=y
\langle\varphi_{45}\rangle$. This degeneracy is broken 
through loop corrections (see third paper in 
Ref.~\cite{other}), but the splitting in our case is tiny.
Higher dimensional operators may induce a larger splitting 
which we will not pursue here.

These neutral fermions interact with the $Z$ boson since they
belong to electroweak doublets. Therefore, they can scatter 
off nucleons by exchanging a $Z$ boson in the $t$-channel. The
spin-independent cross section is \cite{witten}
\beq
\sigma_{\rm SI}=\frac{\mu^2}{\pi}\left(\frac{Zf_p+(A-Z)f_n}{A}
\right)^2,
\eeq  
where $\mu$ is the reduced mass of the DM particle-nucleon 
system, which can be approximated by the proton mass 
$m_{\rm p}$ for $m_{\rm DM}\gg m_{\rm p}$. Also, $Z$ and $A$ 
are the atomic and mass numbers of the nucleus, and
\beq
f_p=\frac{g_{\rm DM}}{m_Z^2}(2g_u+g_d),\quad 
f_n=\frac{g_{\rm DM}}{m_Z^2}(g_u+2g_d).
\eeq
Here $g_{\rm DM}=g_Z/2$, $g_u=(1/2-4\sin^2\theta_{\rm W}/3)
g_Z$, and $g_d=(-1/2+2\sin^2\theta_{\rm W}/3)g_Z$ with $g_Z=
m_Z/\sqrt{2}v$, $v=174~{\rm GeV}$. For $Z=54$ and $A=131$
corresponding to $^{131}$Xe used in the XENON 1T experiment 
\cite{xenon1t}, one finds
\beq
\sigma_{\rm SI}\simeq 2.85\times 10^{-12}~{\rm GeV}^{-2}.
\eeq
The DM in our model can, in general, consist of intermediate 
scale fermions and axions, with the axion fraction given by
$R_{\rm a}=\Omega_{\rm a} h^2/\Omega_{\rm DM}h^2$. Here 
$\Omega_{\rm a} h^2$ is the relic axion abundance and 
$\Omega_{\rm DM}h^2\simeq 0.12$ \cite{planck} is the total 
relic DM abundance. In the presence of axions, the current 
experimental bound on $\sigma_{\rm SI}$ from the XENON 1T 
experiment \cite{xenon1t} can be written as 
\beq
\sigma_{\rm SI}\lesssim \frac{2.21\times 10^{-18}}
{1-R_{\rm a}}\left(
\frac{m_{\rm DM}}{1~{\rm TeV}}\right)~{\rm GeV}^{-2},
\eeq
which implies that
\beq
m_{\rm DM}\gtrsim 1.29\times 10^9\,(1-R_{\rm a})~{\rm GeV}.
\label{DD}
\eeq 
We see that the XENON 1T constraint on the SI cross section
of DM scattering off nuclei requires that the mass of the 
DM particles is at least of intermediate scale exceeding 
the reheat temperature in Eq.~(\ref{reheat}). Consequently, 
thermal DM is excluded and we are led to consider 
non-thermal production of DM particles with at least 
intermediate scale masses via the inflaton decay. (Non-thermal 
superheavy DM particles, called wimpzillas, were previously 
discussed in Ref.~\cite{wimpzilla}.)

We first estimate the relative number density of these DM
particles 
$Y_{\rm DM}=n_{\rm DM}/s$ 
required to reproduce a fraction $(1-R_{\rm a})$ of the 
present DM abundance $\Omega_{\rm DM} h^2\simeq 0.12$ 
\cite{planck} ($n_{\rm DM}$ is the number density of DM 
particles and $s$ is the entropy density) using the 
relation
\beq
(1-R_{\rm a})\,\Omega_{\rm DM} h^2=\frac{m_{\rm DM}
Y_{\rm DM}s_0}{\rho_{\rm c}},
\label{eq:DMY}
\eeq   
where $s_0\simeq 2890~{\rm cm}^{-3}$ is the present entropy 
density and $\rho_{\rm c}\simeq 1.05\times 10^{-5}~{\rm GeV}
\,{\rm cm}^{-3}$ is the present critical density. We obtain
\beq
m_{\rm DM}Y_{\rm DM}\simeq 4.36\times 10^{-10}\,(1-R_{\rm a})
~{\rm GeV}.
\label{mYDM}
\eeq
From the energy density of the DM fermions at reheating, 
$\rho_{\rm DM}=m_{\rm DM}Y_{\rm DM}s(T_{\rm r})$, we then 
find that at $T_{\rm r}$,
\beq
\frac{\rho_{\rm DM}}{\rho_{\rm r}}\simeq \frac{5.81\times 
10^{-10}}{T_{\rm r}}(1-R_{\rm a}),
\eeq
where $\rho_{\rm r}$ is the radiation energy density.
Assuming that the total energy of the inflaton at reheating
is transferred to $\rho_{\rm DM}$ and $\rho_{\rm r}$, the 
inflaton decay width $\Gamma_{\phi\to{\rm DM}}$  
to a pair of DM fermions should satisfy the requirement
$\Gamma_{\phi\to{\rm DM}}/\Gamma_{\phi\to\nu^c}\simeq \rho_{\rm DM}/
\rho_{\rm r}\simeq 1.4 \times 10^{-17}\,(1-R_{\rm a})$.
This yields 
\begin{equation}
\Gamma_{\phi\to{\rm DM}}\simeq 6.78 \times 10^{-20}\,(1-R_{\rm a})
~{\rm GeV}.
\label{GDM}
\end{equation}
We should now check whether our model can reproduce this 
$\Gamma_{\phi\to{\rm DM}}$.
 
\begin{figure}[t]
\centerline{\epsfig{file=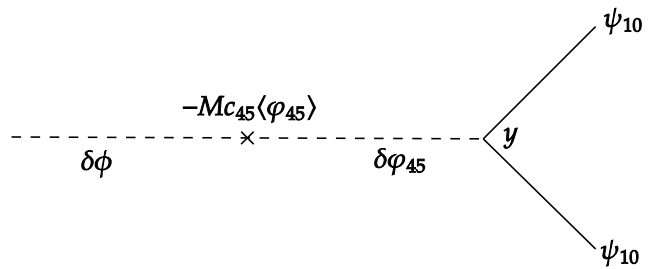,width=8.4cm}}
\caption{Diagram for the inflaton decay into DM (doublets in 
the two 10-plets). The conventions are as in 
Fig.~\ref{fig:infdecaytonuc}.}
\label{fig:infdecaytoDM}
\end{figure}

The diagram for the inflaton decay to a pair of
DM fermions is given in Fig.~\ref{fig:infdecaytoDM}.
The cross sign represents the coefficient $-Mc_{45}
\langle\varphi_{45}\rangle$ of the bilinear coupling 
between $\delta\phi$ and $\delta\varphi_{45}$ (see 
Eq.~(\ref{bilinear})), and $y$ is the Yukawa coupling
constant of $\varphi_{45}$ with the two $\psi_{10}$'s. 
The DM fermions are the neutral components of the 
$SU(2)_L$ doublets in the two $\psi_{10}$'s. The 
propagating $\delta\varphi_{45}$ as well as 
$\langle\varphi_{45}\rangle$ are along the (1,1,3) 
component of $\varphi_{45}$. The inflaton also decays 
into a pair of charged fermions contained in the 
$SU(2)_L$ doublets via the diagram of 
Fig.~\ref{fig:infdecaytoDM}, and into a color triplet 
and an anti-triplet contained in the $\psi_{10}$'s via 
a similar diagram where
the propagating $\delta\varphi_{45}$ and the $\langle
\varphi_{45}\rangle$ are taken along the (15,1,1)
component of $\varphi_{45}$. The VEV of the (15,1,1)
component has to be somewhat larger than the VEV of
the (1,1,3) component in order for the color 
(anti-)triplets to be heavier than the doublets and be 
able to decay into them. However, the required mass 
difference is
much smaller than the (anti-)triplet mass since the 
decay byproducts are SM particles. For simplicity, we 
will assume that the VEVs along the (1,1,3) and 
(15,1,1) components are about equal. As mentioned, the 
color (anti-)triplet and the charged fermions
eventually decay into neutral DM fermions. Therefore,
the inflaton decays either into two neutral, or two charged, 
or two color triplet Dirac fermions with about the same 
width, and all these particles yield neutral fermions 
contributing to DM. The decay width of the inflaton into 
a pair of Dirac fermions should then be multiplied by 
$\approx 10$ to obtain the total decay width.          

For $m_\phi\geq m_{45}$ (see below), the $\delta
\varphi_{45}$ propagator is dominated by the inflaton 
mass $m_\phi$ and the total decay width of the inflaton 
to a pair of DM fermions is given by 
\begin{gather}
\Gamma_{\phi\to{\rm DM}}\simeq\frac{10}{16\pi}
\left(\frac{Mc_{45}\langle\varphi_{45}\rangle y}
{m_\phi^2}\right)^2 m_\phi=
\nonumber\\
\frac{10}{16\pi}\left(\frac{m_{45}^2 m_{\rm DM}}
{Mm_\phi^2}\right)^2 m_\phi,
\label{GDMformula}
\end{gather}
where $m_{\rm DM}=y\langle\varphi_{45}\rangle$. 
Eq.~(\ref{GDM}) then yields
\begin{equation}
m_{45}^2 m_{\rm DM}\simeq 2.96\times 10^{30}\,(1-R_{\rm a})
^{\frac{1}{2}}~{\rm GeV}^3,
\label{45210}
\end{equation}
where $m_{45}$ is the Higgs mass in the (1,1,3) direction of
$\varphi_{45}$ (for simplicity we ignore the mixing between 
the (1,1,3) and (15,1,1) components).  

The relic axion abundance is given \cite{gondolo} by
\begin{equation}
\Omega_{\rm a} h^2\simeq 0.236\left(\frac{f_{\rm a}}{10^{12}
~{\rm GeV}}\right)^{\frac{7}{6}}\langle
{\theta^2f(\theta)}\rangle,
\label{omega}
\end{equation}
where $f_{\rm a}$ is the axion decay constant, i.e. the total
VEV of $\varphi_{45}$, which is greater than $\langle
\varphi_{45}\rangle$, its VEV along its (1,1,3) component 
(see below).  
The misalignment angle $\theta$ lies \cite{curvaton} in the 
interval $[-\pi,+\pi]$ since, in our case, $N$, the sum 
of the $Q_{\rm PQ}^{\prime\prime}$ charges of all fermion
color triplets and anti-triplets, is equal to unity.  All 
$\theta$'s in this 
interval are equally probable. The function $f(\theta)$ 
accounts for the anharmonicity of the axion potential, and 
the average $\langle{\theta^2f(\theta)}\rangle$ is evaluated 
in the above interval and turns out to be \cite{gondolo} 
about 8.77. For definiteness, we take $\alpha_{45}=1/2$, 
which implies $m_{45}=\langle\varphi_{45}\rangle$. We then 
substitute $R_{\rm a}$ in Eq.~(\ref{45210}) by using 
Eq.~(\ref{omega}) and solve the resulting equation to find 
$m_{45}=\langle\varphi_{45}\rangle$ for given values 
of $m_{\rm DM}$ and $f_{\rm a}$. Recall that the VEV of
$\varphi_{45}$ along the (15,1,1) component should be 
somewhat greater than its VEV along the (1,1,3) component, 
so that the color (anti-)triplets in $\psi_{10}$ can decay 
into the $SU(2)_L$ doublets. Consequently, the axion decay 
constant $f_{\rm a}\gtrsim\sqrt{2}\langle\varphi_{45}
\rangle$. For definiteness, we have chosen these VEVs to 
be about equal, which fixes $f_{\rm a}$ close to 
$\sqrt{2}\langle\varphi_{45}\rangle$. Then, for $y=1$, 
we obtain 
\begin{gather}
m_{45}=\langle\varphi_{45}\rangle=m_{\rm DM}\simeq 
1.4\times 10^{10}~{\rm GeV},
\nonumber\\
 f_{\rm a}\simeq 2 \times 10^{10}~{\rm GeV}.
\label{massDM}
\end{gather}
DM is composed of $17.6\%$ axions and $82.4\%$ intermediate 
scale fermions. 
For $m_{\rm DM}=
3\times 10^{9}~{\rm GeV}$, we find 
\begin{gather}
m_{45}=\langle\varphi_{45}\rangle\simeq 2.76\times 
10^{10}~{\rm GeV},\quad f_{\rm a}\simeq 3.9\times 10^{10}~
{\rm GeV},
\nonumber\\
y\simeq 0.11.
\end{gather} 
DM is made up of $39.2\%$ axions and $60.8\%$ intermediate 
scale fermions. 
For $m_{\rm DM}=10^{9}~{\rm GeV}$, we find 
\begin{gather}
m_{45}=\langle\varphi_{45}\rangle\simeq 4.2\times 
10^{10}~{\rm GeV},\quad f_{\rm a}\simeq 6\times 10^{10}~
{\rm GeV},
\nonumber\\ 
y\simeq 2.38\times 10^{-2}.
\end{gather} 
DM consists of $63.9\%$ axions and $36.1\%$ intermediate 
scale fermions. 
Finally, for $m_{\rm DM}=3\times 10^{8}~{\rm GeV}$, we 
find 
\begin{gather}
m_{45}=\langle\varphi_{45}\rangle\simeq 5.6\times 
10^{10}~{\rm GeV},\quad f_{\rm a}\simeq 7.93\times 
10^{10}~{\rm GeV}, 
\nonumber\\
y\simeq 5.35\times 10^{-3}.
\end{gather} 
$89.6\%$ of DM consists of axions and $10.4\%$ of 
intermediate scale fermions. These values of 
$m_{45}$ clearly satisfy the requirement that the 
inflaton decay into a pair of electroweak Higgs fields 
and a SM singlet scalar (see Fig.~\ref{fig:infdecaytoh}) 
is subdominant. Also, the values of $m_{\rm DM}$ satisfy 
the requirement from direct detection of DM in 
Eq.~(\ref{DD}) and the kinematic constraint $m_\phi\geq 
2m_{\rm DM}$ which makes the decay possible. Moreover, 
$m_{\rm DM}$ exceeds the reheat temperature and $m_{45}$ 
is smaller than $m_\phi$, consistent with our assumption 
in deriving Eq.~(\ref{GDMformula}). We can also 
differentiate $m_{45}$ and $\langle\varphi_{45}\rangle$ 
by taking $\alpha_{45}\neq 1/2$, so as to increase 
$\langle\varphi_{45}\rangle$ and, thus, the axion decay 
constant $f_{\rm a}$.

\begin{figure}[t]
\centerline{\epsfig{file=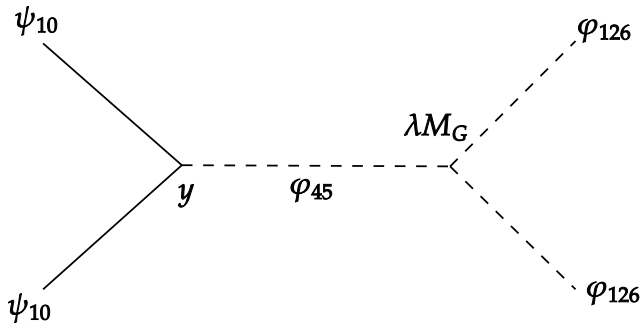,width=8.4cm}}
\caption{Diagram for DM pair annihilation to two 
electroweak Higgs fields. The conventions are as in 
Fig.~\ref{fig:infdecaytoDM}.}
\label{fig:DMannih}
\end{figure}

In order to complete the DM discussion, we have to show 
that the pair annihilation of DM fermions is out of
equilibrium at all temperatures smaller than the reheat
temperature so that their abundance remains constant. A 
dominant diagram for this annihilation is shown in 
Fig.~\ref{fig:DMannih}. The propagating $\varphi_{45}$ 
lies along its (1,1,3) direction and the decay products 
are a pair of electroweak Higgs fields contained in 
$\varphi_{126}$. The cross section is estimated to be
\beq
\sigma_{\rm DM}\simeq\frac{1}{16\pi^2}\left(\frac{y\lambda 
M_{\rm G}}{4m_{\rm DM}^2}\right)^2.
\label{sigmaDM}
\eeq
The out-of-equilibrium condition reads as follows:
\beq
n_{\rm DM}\sigma_{\rm DM}\lesssim H,
\label{out}
\eeq 
for all $T\lesssim T_{\rm r}$ ($H$ is the Hubble parameter). 
From the Friedmann equation, we find 
\beq
H=\frac{\rho_{\rm r}^{1/2}}{\sqrt{3}\,m_{\rm P}}=\frac{\pi 
g_*^{1/2}T^2}{3\sqrt{10}\,m_{\rm P}},
\label{Hubble}
\eeq
and Eq.~(\ref{mYDM}) gives
\begin{gather}
n_{\rm DM}=sY_{\rm DM}=
\nonumber\\
\frac{2\pi^2g_*T^3}{45}\,(1-R_{\rm a})
\left(\frac{m_{\rm DM}}{4.36\times 10^{-10}~{\rm GeV}}
\right)^{-1}. 
\label{nDM}
\end{gather}
For the case in Eq.~(\ref{massDM}) and using 
Eqs.~(\ref{sigmaDM}), 
(\ref{Hubble}), and (\ref{nDM}) with the maximal 
allowed value of $\lambda$ which is of order 
$m_{45}/M_{\rm G}$, 
we see that the condition in Eq.~(\ref{out}) is
very well satisfied for all $T\lesssim T_{\rm r}$.
We conclude that the pair annihilation of DM fermions
in Fig.~\ref{fig:DMannih} 
is utterly suppressed at all relevant temperatures. We 
should note that one could instead consider the DM pair
annihilation into SM particles via the exchange
of a $Z$ boson in the $s$-channel. The corresponding
cross section is of the same order of magnitude as the
cross section in Eq.~(\ref{sigmaDM}) and
our conclusion therefore would be the same. 

The spontaneous breaking of the PQ symmetry at a scale
of order $10^{10}-10^{11}~{\rm GeV}$ takes place after
the end of inflation. Indeed, in the numerical example 
under consideration, inflation terminates when the 
inflaton field reaches the value $\phi_{\rm e}\simeq 
6.83\times 10^{19}~{\rm GeV}$ \cite{okada}. From 
Eq.~(\ref{CW}), we find that $V(\phi_{\rm e})^{1/4}
\simeq 6.31\times 10^{15}~{\rm GeV}$, yielding the Hubble 
parameter $H_e\simeq 9.42\times 10^{12}~{\rm GeV}$. 
The field $\varphi_{\theta}$ develops a VEV and the 
corresponding phase transition takes place when 
$c_{\theta}\phi^2/2\sim \sigma T_{\rm H}^2$ 
\cite{Shafi:1983bd}, where 
$\sigma\sim 1$ and $T_{\rm H}=H/2\pi$ is the Hawking 
temperature. Consequently, for a phase transition 
which occurs before the end of inflation we must have
$c_{\theta}\gtrsim H_e^2/2\pi^2\phi_{\rm e}^2\simeq 9.63
\times 10^{-16}$. This implies that the corresponding 
scale $\langle\varphi_{\theta}\rangle\gtrsim 2.23
\times 10^{12}~{\rm GeV}$, which excludes the PQ 
transition. Therefore, the presence of the two fermion
10-plets which lead to a solution of the axion domain 
wall problem via the Lazarides-Shafi 
mechanism \cite{axionwalls} is vital. At reheating, the 
masses of the scalar fields $\varphi_\theta$ acquire 
temperature corrections which, however, are subdominant 
compared to the first term in Eq.~(\ref{ctheta}). Indeed, 
the decaying inflaton oscillates about $M$ and, thus, 
$c_{45}=(m_{45}/M)^2\simeq 7.78\times 10^{-20}$ for 
$m_{45}$ in Eq.~(\ref{massDM}). Consequently, $c_{45}
\phi^2/2\gg T_{\rm r}^2$, for $T_{\rm r}$ in 
Eq.~(\ref{reheat}). The PQ symmetry is already broken at 
reheating and the DM fermions have acquired their masses. 

As we previously mentioned, the gauge symmetry breaking 
at the intermediate scale $M_{\rm I}$ generates 
topologically stable $Z_2$ cosmic strings. The 
dimensionless string tension $G\mu_{\rm s}$, where 
$G$ is Newton's gravitational constant and 
$\mu_{\rm s}$ the string tension, i.e. the energy per 
unit length of the string, is given by
\beq
G\mu_{\rm s}\simeq \frac{1}{8} \left(\frac{M_{\rm I}}
{m_{\rm P}}\right)^2.
\eeq
Here we assumed that these strings are close to the 
Bogomol'nyi limit of the Abelian Higgs model 
\cite{bevis}. A recent pulsar timing array $95\%$ 
confidence level limit on the dimensionless string 
tension is \cite{olum}
\beq
G\mu_{\rm s}\lesssim 1.5\times 10^{-11},
\label{Gmu}
\eeq
which holds for strings surviving until the present time.
Eq.~(\ref{Gmu}) implies the following upper bound on the 
intermediate scale
\beq
M_{\rm I}\lesssim 2.67\times 10^{13}~{\rm GeV}.
\eeq
Note that strings corresponding to such intermediate 
scales are not 
inflated away as shown in Ref.~\cite{topdef} and, thus, 
the limit in Eq.~(\ref{Gmu}) applies. These strings are 
possibly measurable by LISA in the future. Applying
the analysis of Ref.~\cite{topdef}, we find that, for the
strings to be inflated away, the number of e-foldings
following their generation should exceed about 68 and,
thus, $c_{126}\gtrsim 1.94\times 10^{-12}$ and 
\beq
M_{\rm I}\gtrsim 10^{14}~{\rm GeV}.
\eeq  
In this case, the cosmic strings are not restricted by 
Eq.~(\ref{Gmu}). The value of the inflaton field 
$\phi_{\rm I}$ at which the intermediate transition 
takes place is found from the relation $c_{126}
\phi_{\rm I}^2/2\sim T_{\rm H}^2$ and, thus,
$\phi_{\rm I}\lesssim 1.2\times 10^{19}~{\rm GeV}
\simeq 4.9\,m_{\rm P}$. The GUT magnetic monopoles are 
certainly inflated away since $M_{\rm G}\gg 10^{14}
~{\rm GeV}$ \cite{topdef}.  

\section{Non-thermal Leptogenesis}
\label{sec:lepto}

The observed BAU $Y_B=n_B/s\simeq 8.69\times 10^{-11}$ 
\cite{planck} can be reproduced in our model via 
non-thermal leptogenesis \cite{nonthermalepto}, i.e. 
the generation of a primordial lepton asymmetry 
$Y_L=n_L/s$ \cite{thermallepto} at reheating which, at 
the electroweak transition, is partially converted
into the observed BAU via sphaleron effects. For 
non-supersymmetric SM, $Y_B\simeq-0.35Y_L$. As we have
already discussed, the inflaton predominantly decays 
into a pair of $\nu^c_2$'s, where $\nu^c_2$ is the 
second heaviest right-handed neutrino with mass 
$M_2\simeq 8.5\times 10^{12}~{\rm GeV}$. (The decay 
into first generation right-handed neutrinos is 
suppressed because of their smaller coupling to the 
inflaton.) The 
primordial lepton asymmetry will be produced 
non-thermally \cite{nonthermalepto,vlachos} via the
subsequent out-of-equilibrium decay of this 
right-handed neutrino into an electroweak Higgs doublet 
and a lepton or anti-lepton via the exchange of the 
heaviest $\nu^c$ with mass $M_3=z^\prime\langle
\varphi_{126}\rangle$. We will not explore the 
feasibility of thermal leptogenesis in this work. The 
relevant one-loop diagrams 
are both of the vertex and self-energy type \cite{covi}.
Recall that $\langle\varphi_{126}\rangle$ can have any
value greater than or equal to $m_\phi$, and thus 
$M_2/M_3$ can be adjusted at any value smaller than 
unity. However, it should not be too small since it will 
suppress the BAU (see below), but also not too close 
to unity since the validity of our calculation requires 
\cite{apostolos} that $M_2\ll M_3$ and $\Gamma_{\phi\to\nu^c}
\ll (M_3^2-M_2^2)/M_2$.

Under these assumptions and considering only the two 
heavier generations, $Y_B$ can be approximated as 
\cite{vlachos}
\begin{equation}
Y_B\simeq 0.35\,\frac{9}{16\pi}\,\frac{T_{\rm r}}
{m_\phi}\,\frac{M_2}{M_3}\frac{{\rm c}^2{\rm s}^2
\sin2\delta(m_{3}^{D}\,^{2}
-m_{2}^{D}\,^{2})^2}{v^2(m_{3}^{D}\,^{2}{\rm s}^2+
m_{2}^{D}\,^{2}{\rm c}^2)}.
\label{YB}
\end{equation}        
Here $v\simeq 174~{\rm GeV}$, ${\rm c}=\cos\theta$, 
${\rm s}=\sin\theta$, with $\theta$ and $\delta$ 
being the rotation angle and phase which diagonalize 
the Majorana mass matrix of $\nu^c$\,'s
in the basis where the Dirac neutrino mass matrix is
diagonal with eigenvalues $m_{2}^{D}$ and $m_{3}^{D}$. 
The determinant and trace invariants of the light 
neutrino mass matrix imply \cite{vlachos} that the 
neutrino parameters should satisfy the following 
constraints:
\begin{equation}
m_{2}m_{3}=\frac{\left(m_{2}^{D}m_{3}^{D}\right)^{2}}
{M_{2}M_{3}}~,
\label{eq:determinant}
\end{equation}
\begin{gather}
m_{2}\,^{2}+m_{3}\,^{2}=\frac{\left(m_{2}^{D}\,^{2}
{\rm \ c}^{2}+m_{3}^{D}\,^{2}{\rm \ s}^{2}\right)^{2}}
{M_{2}\,^{2}}+
\nonumber\\
\frac{\left(m_{3}^{D}\,^{2}{\rm \ c}^{2}+
m_{2}^{D}\,^{2}{\rm \ s}^{2}\right)^{2}}{M_{3}\,^{2}}+
\frac{2(m_{3}^{D}\,^{2}-m_{2}^{D}\,^{2})^{2}
{\rm c}^{2}{\rm s}^{2}{\cos 2\delta }}{M_{2}M_{3}}~.
\label{eq:trace}
\end{gather}
Here we assume a normal hierarchy of light neutrino 
masses $m_i$ ($i=1,2,3$) \cite{valle}, with  
$m_{3}\simeq 5.05\times 10^{-2}~{\rm eV}$, 
$m_{2}\simeq 8.73\times 10^{-3}~{\rm eV}$, 
and $m_1\simeq 0$.
 
For a rough estimate of a possible solution of the system 
of Eqs.~(\ref{YB}), (\ref{eq:determinant}), and 
(\ref{eq:trace}), we take ${\rm c}^2\simeq {\rm s^2}
\simeq 1/2$, $\sin2\delta\simeq 1$. Note that the latter
choice maximizes $Y_B$. Substituting $Y_B$ with its 
observed value in Eq.~(\ref{YB}), we are left with just 
three unknown 
variables $\beta\equiv M_2/M_3$, $m_{2}^{D}$, $m_{3}^{D}$, 
and we can solve the system of these three equations to 
determine them. To this end, we find from  
Eq.~(\ref{eq:trace}) that
\beq
m_{2}^{D}\,^{2}+m_{3}^{D}\,^{2}\simeq 876.36
\left(1+\beta^2\right)^{-1/2}~{\rm GeV}^2,
\label{sum}
\eeq
while Eq.~(\ref{eq:determinant}) gives
\begin{gather}
\left(m_{2}^{D}\,^{2}-m_{3}^{D}\,^{2}\right)^2\simeq
\nonumber\\
10^5\left(7.68(1+\beta^2)^{-1}-0.645 \beta^{-1}
\right)~{\rm GeV}^4.
\label{dif}
\end{gather}
Substituting these two equations in Eq.~(\ref{YB}), we 
obtain
\begin{equation}
0.303\simeq 
7.68\,\beta(1+\beta^2)^{-1/2}-0.645
(1+\beta^2)^{1/2},
\end{equation}
which is solved numerically and yields $\beta\simeq 0.125$.
This implies that 
\begin{equation}
M_3\simeq 6.84\times 10^{13}~{\rm GeV}.
\end{equation} 
From Eqs.~(\ref{sum}) and (\ref{dif}), we estimate the
Dirac neutrino masses:
\beq
m_{2}^{D}\simeq 14~
{\rm GeV},\quad m_{3}^{D}\simeq 26.2~{\rm GeV}.
\eeq
Clearly, this is just an example to show that the 
observed BAU can be generated in our model in accord 
with the neutrino experimental data. A more complete 
and accurate calculation including all three 
generations of neutrinos should be carried out. In 
any case, since the neutrino Dirac mass matrix has a 
certain degree of freedom, we believe that more realistic
solutions can be found. Note that the requirements
$M_2\ll M_3$ and $\Gamma_{\phi\to\nu^c}\ll (M_3^2-M_2^2)/M_2$ 
are well satisfied. Also $M_2\gg T_{\rm r}$ and so
the second heaviest $\nu^c$ decays out of equilibrium 
to generate the primordial lepton asymmetry. Finally, note
that right-handed neutrinos should all be heavier than 
$T_{\rm r}$ to prevent the erasure of the lepton asymmetry.
With such low reheat temperature any discussion of thermal 
leptogenesis is beyond the scope of this paper.

\section{Conclusions}
\label{sec:concl}

We have explored some interesting predictions of a  
non-supersymmetric $SO(10)\times U(1)_{\rm PQ}$ model in 
which the spontaneous breaking of $U(1)_{\rm PQ}$ takes 
place after inflation. A pair of 10-plet fermions with 
intermediate scale masses comparable to or somewhat 
smaller than the axion decay constant $f_{\rm a}$ are 
introduced in order to evade the 
axion domain wall problem. The electroweak doublets from 
these 10-plets provide a novel non-thermal dark matter 
candidate whose stability is guaranteed by an unbroken 
$Z_2$ symmetry. We discuss an explicit realization of this 
scenario by incorporating inflation driven by an $SO(10)
\times U(1)_{\rm PQ}$ singlet scalar field with a 
Coleman-Weinberg potential. The dark matter fermions have 
mass on the order of $3\times 10^{8}-10^{10}~{\rm GeV}$ 
and, in addition, non-thermal leptogenesis is realized. 
The model also yields topologically stable intermediate 
mass scale cosmic strings which survive inflation and emit 
\cite{topdef} possibly observable gravity waves (for a 
recent discussion in supersymmetric $SO(10)$, see 
Ref.~\cite{kai}. Last, but not least, the tensor-to-scalar 
ratio $r$, a canonical measure of gravity waves generated 
during inflation, cannot be smaller than 0.01 and therefore 
should be accessible in the next generation experiments.

\vspace{.5cm}

\noindent 
{\bf Acknowledgments.}\,
{Q.S thanks Nobuchika Okada for very helpful discussions and 
clarifications related to thermal and non-thermal dark matter.
G.L. thanks John Vergados for discussions on dark matter 
detection. This work is supported in part by the DOE 
Grant DE-SC-001380. It is also supported by the Hellenic 
Foundation for Research and Innovation (H.F.R.I.) under the 
``First Call for H.F.R.I. Research Projects to support 
Faculty Members and Researchers and the procurement of 
high-cost research equipment grant'' (Project Number: 2251).
We thank Amit Tiwari for his help with the figures.}

\def\ijmp#1#2#3{{Int. Jour. Mod. Phys.}
{\bf #1},~#3~(#2)}
\def\plb#1#2#3{{Phys. Lett. B }{\bf #1},~#3~(#2)}
\def\zpc#1#2#3{{Z. Phys. C }{\bf #1},~#3~(#2)}
\def\prl#1#2#3{{Phys. Rev. Lett.}
{\bf #1},~#3~(#2)}
\def\rmp#1#2#3{{Rev. Mod. Phys.}
{\bf #1},~#3~(#2)}
\def\prep#1#2#3{{Phys. Rep. }{\bf #1},~#3~(#2)}
\def\prd#1#2#3{{Phys. Rev. D }{\bf #1},~#3~(#2)}
\def\npb#1#2#3{{Nucl. Phys. }{\bf B#1},~#3~(#2)}
\def\np#1#2#3{{Nucl. Phys. B }{\bf #1},~#3~(#2)}
\def\npps#1#2#3{{Nucl. Phys. B (Proc. Sup.)}
{\bf #1},~#3~(#2)}
\def\mpl#1#2#3{{Mod. Phys. Lett.}
{\bf #1},~#3~(#2)}
\def\arnps#1#2#3{{Annu. Rev. Nucl. Part. Sci.}
{\bf #1},~#3~(#2)}
\def\sjnp#1#2#3{{Sov. J. Nucl. Phys.}
{\bf #1},~#3~(#2)}
\def\jetp#1#2#3{{JETP Lett. }{\bf #1},~#3~(#2)}
\def\app#1#2#3{{Acta Phys. Polon.}
{\bf #1},~#3~(#2)}
\def\rnc#1#2#3{{Riv. Nuovo Cim.}
{\bf #1},~#3~(#2)}
\def\ap#1#2#3{{Ann. Phys. }{\bf #1},~#3~(#2)}
\def\ptp#1#2#3{{Prog. Theor. Phys.}
{\bf #1},~#3~(#2)}
\def\apjl#1#2#3{{Astrophys. J. Lett.}
{\bf #1},~#3~(#2)}
\def\apjs#1#2#3{{Astrophys. J. Suppl.}
{\bf #1},~#3~(#2)}
\def\n#1#2#3{{Nature }{\bf #1},~#3~(#2)}
\def\apj#1#2#3{{Astrophys. J.}
{\bf #1},~#3~(#2)}
\def\anj#1#2#3{{Astron. J. }{\bf #1},~#3~(#2)}
\def\mnras#1#2#3{{MNRAS }{\bf #1},~#3~(#2)}
\def\grg#1#2#3{{Gen. Rel. Grav.}
{\bf #1},~#3~(#2)}
\def\s#1#2#3{{Science }{\bf #1},~#3~(#2)}
\def\baas#1#2#3{{Bull. Am. Astron. Soc.}
{\bf #1},~#3~(#2)}
\def\ibid#1#2#3{{\it ibid. }{\bf #1},~#3~(#2)}
\def\cpc#1#2#3{{Comput. Phys. Commun.}
{\bf #1},~#3~(#2)}
\def\astp#1#2#3{{Astropart. Phys.}
{\bf #1},~#3~(#2)}
\def\epjc#1#2#3{{Eur. Phys. J. C}
{\bf #1},~#3~(#2)}
\def\nima#1#2#3{{Nucl. Instrum. Meth. A}
{\bf #1},~#3~(#2)}
\def\jhep#1#2#3{{J. High Energy Phys.}
{\bf #1},~#3~(#2)}
\def\jcap#1#2#3{{J. Cosmol. Astropart. Phys.}
{\bf #1},~#3~(#2)}
\def\lnp#1#2#3{{Lect. Notes Phys.}
{\bf #1},~#3~(#2)}
\def\jpcs#1#2#3{{J. Phys. Conf. Ser.}
{\bf #1},~#3~(#2)}
\def\aap#1#2#3{{Astron. Astrophys.}
{\bf #1},~#3~(#2)}
\def\mpla#1#2#3{{Mod. Phys. Lett. A}
{\bf #1},~#3~(#2)}

\end{document}